\documentclass[12pt,preprint]{aastex}

\usepackage{graphicx}
\usepackage{rotating}

\newcommand{\appropto}{\mathrel{\vcenter{
  \offinterlineskip\halign{\hfil$##$\cr
\propto\cr\noalign{\kern2pt}\sim\cr\noalign{\kern-2pt}}}}}

\shorttitle{}
\shortauthors{Nesvorn\'y et al.}
\begin{document}
\baselineskip 19.pt

\title{Impact Rates in the Outer Solar System}

\author{David Nesvorn\'y, Luke Dones}
\affil{Department of Space Studies, Southwest Research Institute,
  1050 Walnut St., Suite 300, Boulder, CO 80302, USA}
\author{Mario De Pr\'a}
\affil{Florida Space Institute, University of Central Florida,
12354 Research Parkway, Partnership 1 Building, Suite 214,
Orlando, FL 32826-0650}
\author{Maria Womack}
\affil{National Science Foundation, 2415 Eisenhower Avenue,
  Alexandria, VA 22314}
\affil{Department of Physics, University of Central Florida, 4000
  Central Florida Boulevard, Building 12, 310, Orlando, FL 32816}

\author{Kevin J. Zahnle}
\affil{NASA Ames Research Center, MS 245-3, Moffett Field, CA 94035}

\begin{abstract}
Previous studies of cometary impacts in the outer Solar System used
the spatial distribution of ecliptic comets (ECs) from dynamical
models that assumed ECs began on low-inclination orbits ($\lesssim
5^\circ$) in the Kuiper belt \citep{Levison1997}. In reality, the
source population of ECs -- the trans-Neptunian scattered disk -- has
orbital inclinations reaching up to $\sim 30^\circ$
\citep{DiSisto2020}.  In \citet{Nesvorny2017}, we developed a new
dynamical model of ECs by following comets as they evolved from the
scattered disk to the inner Solar System. The model was absolutely
calibrated from the population of Centaurs \citep{Nesvorny2019} and
active ECs. Here we use our EC model to determine the steady-state
impact flux of cometary/Centaur impactors on Jupiter, Saturn, Uranus,
and their moons. Relative to previous work \citep{Zahnle2003}, we find
  slightly higher impact probabilities on the outer moons and lower
  impact probabilities on the inner moons. The impact probabilities
  are smaller when comet disruption is accounted for.  The results
  provide a modern framework for the interpretation of the cratering
  record in the outer Solar System.
\end{abstract}

\section{Introduction}

Comets are icy objects that reach the inner Solar System after leaving
distant reservoirs beyond Neptune and are dynamically evolving onto
elongated orbits with small perihelion distances (see
\citet{Dones2015} and \citet{Kaib2023} for recent reviews). Their
activity, manifesting itself by the presence of a dust/gas coma and
tail, is driven largely by solar heating and sublimation of various
ices. Once they reach the inner Solar System, comets are short-lived,
implying that they must be resupplied from external reservoirs. Here
we focus on the ecliptic comets (ECs -- low-to-moderate inclination
planet-crossing bodies in the region of the giant planets; see
\citet{Levison1997} and \citet{Fraser2023} for
further discussion) because (1) the population of ECs is relatively
well characterized from observations and allows us to construct a
realistic model (Section~2), and (2) ECs, and their Centaur
precursors, are more important for impact cratering in the outer Solar
System than other types of comets (e.g., long-period comets; LPCs), 
escaped asteroids and Trojans (e.g., \citet{Levison2000}, 
\citet{Zahnle1998}, \citet{Zahnle2001}, and \citet{Zahnle2003} [hereafter Z03]).

\citet{Levison1997} [hereafter LD97] considered the origin and
evolution of ecliptic comets.  LD97 assumed that the {\it classical} Kuiper belt
30--50 au from the Sun was the main source of ECs. They showed that
small Kuiper belt objects (KBOs) reaching Neptune-crossing orbits can
be scattered by encounters with the outer planets to small perihelion
distances ($q<2.5$~au), at which point they are assumed to become
active and visible Jupiter-family comets (JFCs). The JFCs reaching
$q<2.5$ au for the first time have a narrow inclination distribution
in the LD97 model, because the orbits were assumed to start with low
inclinations in the Kuiper belt and the inclinations changed little
before the comets reached Jupiter-crossing
orbits.\footnote{\citet{Levison2006} assumed a broader inclination
distribution in their study of the origin of comet 2P/Encke, but still
took the classical Kuiper belt to be the source of ecliptic comets.}
\citet{Levison1994} and LD97 pointed out that the inclination distribution of JFCs
widens over time due to scattering encounters with Jupiter. LD97 found
their best fit to the observed inclination distribution of JFCs
(median $\simeq 13^\circ$) when they assumed that ECs remain active
for $\simeq12$,000 years (with a range of 3,000 to 30,000 years) after
first reaching $q<2.5$ au.

The escape of ECs from the classical Kuiper belt is driven by slow
chaotic processes in various orbital resonances with Neptune. Because
these processes affect only part of the belt, with most orbits being
stable for billions of years, comet delivery from the classical Kuiper
belt is inefficient (Nesvorn\'y et al. 2017). \citet{Duncan1997} suggested that the {\it
  scattered disk}\footnote{\citet{Gladman2008} divide the scattered
disk into two populations -- the scattering and detached disks. The
scattering disk is the population of objects currently scattering from
Neptune. The detached disk contains scattered disk objects with larger
perihelion distances that are not interacting strongly with Neptune.},
whose prototype is (15874) 1996~TL66 (semi-major axis $a = 134$~au, $q
= 35$~au; \citet{Luu1997}), should be a more prolific source of ECs.
This is because scattered disk objects (SDOs) can approach Neptune
during their perihelion passages and be scattered by Neptune to orbits
with shorter orbital periods, implying a faster loss rate for SDOs
than for classical KBOs \citep{Duncan2004}. The population of SDOs is
also inferred to be larger than that of classical KBOs ($\sim 0.05$
$M_{\Earth}$ vs. $\sim 0.01$ $M_{\Earth}$; \citet{Fraser2014},
\citet{Nesvorny2018a}) thus representing a large source for ECs.

Z03 used the spatial distribution of ECs from LD97 to determine impact
rates in the outer Solar System.  They estimated that the current rate
at which EC nuclei with diameters $D>1.5$~km strike Jupiter is
$0.005^{+0.006}_{-0.003}$ yr$^{-1}$.\footnote{Z03 took 1.5~km as their
reference diameter based on the \citet{Scotti1993} and
\citet{Asphaug1996} estimates of the size of Shoemaker-Levy 9's
nucleus before Jupiter tidally disrupted the comet in 1992.}  Z03's
rate for Jupiter was based on considerations such as four known
passages of JFCs within 3 jovian radii in the past 150 years and
models of the population of near-Earth JFCs from LD97. Z03 then
calculated impact rates on Saturn, Uranus, and Neptune by scaling from
the rate on Jupiter using LD97's model. Impact rates on the satellites
of the giant planets were, in turn, calculated by Z03 with an
\"Opik-like formalism \citep{Opik1951}.\footnote{The \"Opik method  
is a useful tool for computing the average impact rate between two bodies. 
The method assumes that the orbital longitudes of the two bodies are 
randomly distributed between 0 and $2\pi$ and evaluates the impact 
rate by accounting for all possible intersections of the two orbits.}
The main uncertainties in the
impact rates given in Z03 arise from: (1) the assumed source reservoir
in LD97, and hence the spatial distribution of ECs (see discussion
above); (2) our poor understanding of the sizes of the nuclei of
active ECs; (3) the approximate nature of the impact flux calculation
in LD97 (see \citet{Levison2000}); and (4) the small number of direct
impacts recorded on each planet in LD97.  As for (2), a common
procedure is to infer the diameter $D$ of a comet nucleus from its
absolute magnitude $H_{\rm T}$, which quantifies the total brightness
of a comet, including coma, at a standard distance of 1~au from the
Earth and Sun (e.g., \citet{Brasser2013}). The nature of the relation
between $H_{\rm T}$ and $D$ is, however, uncertain because comets vary
greatly in their levels of activity\footnote{The brightness may also
be affected by scattering and phase angle effects in the coma
\citep{Schleicher2011, Womack2021}.}.

\begin{figure}
\epsscale{0.7}
%\plotone{jfc.eps}
%\plotone{fig1.eps}
\plotone{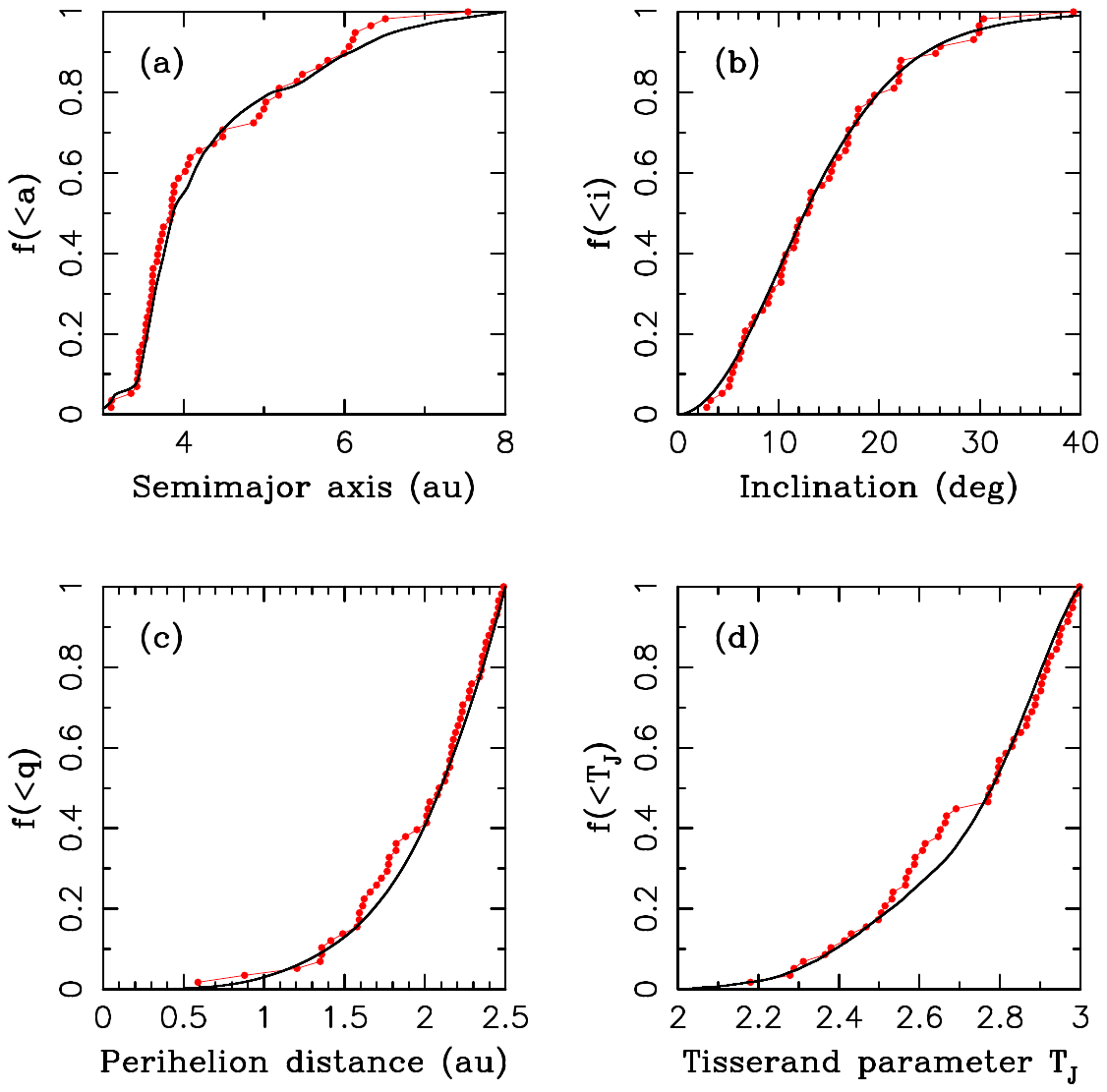}
\caption{The cumulative orbital distributions of ecliptic comets with
  orbital periods $P<20$~yr, Tisserand parameters $2<T_{\rm J}<3$, and
  perihelion distances $q<2.5$ au. The model results (solid lines) are
  compared with the distribution of known ECs (connected red dots). In
  the model, we assumed that visible ECs remain active for 600
  perihelion passages with $q<2.5$ au. The Kolmogorov-Smirnov test for 
  orbital inclinations, which are the main focus here, gives a 91\% probability 
  that the model and observed distributions were drawn from the same 
  distribution. Adapted from N17.}
\label{jfc}
\end{figure}

An important prerequisite for the interpretation of the cratering
record in the outer Solar System is to have a reliable model of
ECs\footnote{We do not discuss the scaling from impactor diameter to
crater size in this work. See, e.g., \citet{Kraus2011},
\citet{Wong2021}, \citet{Holsapple2022}, and \citet{Bottke2023} for
recent treatments of this topic.}.  We developed such a model in
\citet{Nesvorny2017} (hereafter N17). In N17, we performed end-to-end
simulations in which cometary reservoirs, including the scattered disk
and Oort cloud, were produced in the early Solar System and evolved
for 4.5 Gyr. The simulations included the effects of the four giant
planets.  The model was calibrated from the observed population of
active ECs, but includes other types of comets as well. We considered
different scenarios for the duration of cometary activity, including a
simple model in which comets remain active for $N_{\rm p}(q)$
perihelion passages with perihelion distance $q<2.5$ au. To constrain
$N_{\rm p}(2.5)$, we compared the orbital distribution and number of
active comets produced in the model to observations. The observed
distribution was well reproduced with $N_{\rm p}(2.5)\simeq300$--800
(see Fig.~\ref{jfc} here). Here we adopt $N_{\rm p}(2.5)\simeq600$ as a
reference value for comets with $D \sim 1$~km (N17 inferred that
$N_{\rm p}(2.5)$ should increase with the size of the nucleus). As the
median orbital period of ECs is $\simeq 8$ yr (N17), our reference
value corresponds to $\simeq 4$,800 yr -- a factor of 2.5 shorter than
the nominal LD97 estimate. Ultimately, this is a consequence of new
ECs (i.e., bodies reaching Jupiter-crossing orbit for the first time)
having a wider inclination distribution in the N17 model because they
start with larger inclinations in the scattered disk.

The main uncertainty in the N17 model was related to the absolute
calibration. N17 used Jupiter Trojans, whose size distribution is well
characterized from observations down to $D\sim3$ km \citep{Wong2015,
  Yoshida2017}, for this purpose. This method relies on the assumption
that the Trojan implantation efficiency from the original planetesimal
disk is relatively well determined \citep{Morbidelli2005,
  Nesvorny2013}. We modeled the collisional evolution of Jupiter
Trojans and SDOs and found that the size distribution changes {\it
  after} their implantation were insignificant \citep{Nesvorny2018b,
  Bottke2023}.  In addition, Pluto/Charon craters indicate that the
size distribution slope of impactors with $D>1$ km in the Kuiper belt
is similar to that of Jupiter Trojans \citep{Singer2019}.  The use of
the Trojan size distribution to model comets may therefore be
justified.

\section{Method}

\begin{figure}
\epsscale{0.7}
%\plotone{fig2.eps}
\plotone{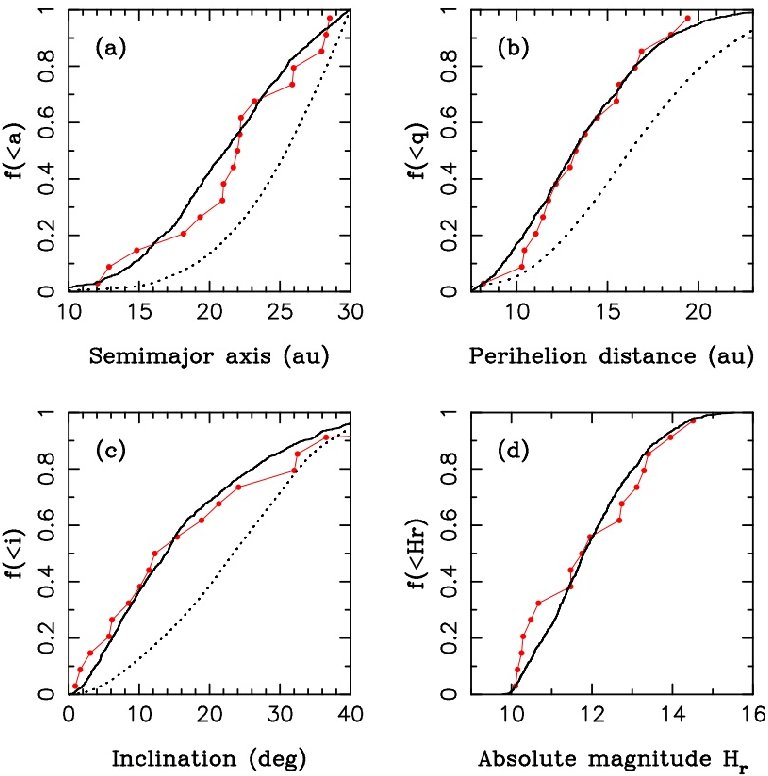}
%\plotone{centaur1.eps}
\caption{The cumulative orbital and magnitude distributions of Centaurs. The 
  biased model results (solid lines) are compared with OSSOS detections of 
  Centaurs (connected red dots).  The Kolmogorov-Smirnov tests give 52\%, 93\%, 
  95\%, and 60\% probabilities for the semimajor axis, perihelion distance, inclination, 
  and magnitude distributions. The differences between the model and 
  observed distributions arise from statistical fluctuations (OSSOS 
  only detected $\simeq 20$ Centaurs). The intrinsic model distributions 
  are shown as dotted lines. Adapted from Nesvorn\'y et al. (2019).}
\label{centaur}
\end{figure}

A better calibration of planetary impactors in the outer Solar System
is provided by the Outer Solar System Origins Survey (OSSOS;
\citet{Bannister2018}) observations of Centaurs (Dorsey et al.\ 2023,
Cabral et al.\ 2019). In
\citet{Nesvorny2019}, we used the N17 model to predict the orbital
distribution and the number of Centaurs. The model distribution was
biased by the OSSOS simulator and compared with the OSSOS Centaur
detections. We found a good fit to the observed orbital distribution,
including the wide range of orbital inclinations, which was the
hardest characteristic to fit in previous models (see Fig.~\ref{centaur} 
here and Marsset et al. 2019 for how the orbital inclinations of KBOs 
and Centaurs correlate with photometric color, and Nesvorn\'y et al.
2020 for modeling of the inclination-color relationship).
The N17 dynamical model, in which the original population of outer
disk planetesimals was calibrated from Jupiter Trojans (see above),
implies that OSSOS should have detected $11 \pm 4$ Centaurs with
semimajor axes $a < 30$ au, perihelion distances $q > 7.5$ au, and
diameter $D > 10$ km (absolute magnitude $H_r < 13.7$ for a 6\%
geometric albedo). This is consistent with 15 actual OSSOS Centaur
detections with $H_r < 13.7$.

By slightly adjusting the N17 model to accurately match the OSSOS
Centaur detections, \citet{Nesvorny2019} infer that the inner
scattered disk at $50 < a < 200$ au, from where most ECs evolve (N17),
should contain $(2.0 \pm 0.8) \times 10^7$ $D > 10$ km objects at the
present time. This is consistent with independent estimates by
\citet{Lawler2018} and \citet{DiSisto2020}.  We can then rewind the
history of the population using the N17 simulations to estimate that
the original trans-Neptunian disk contained $(8 \pm 3) \times 10^9$
planetesimals with $D>10$ km. Reference population estimates for
smaller diameter cutoffs can be obtained from the size distribution of
Jupiter Trojans: cumulative $N(D) \propto D^{-2.1}$ down to $D \sim$
3--5~km (\citet{Grav2011, Wong2015, Yoshida2017, Uehata2022}; see
discussion in Section 3.1), and perhaps even down to $D \sim$~1--2~km.

With the orbital model and absolute calibration in place, we can
estimate the impact rates of ECs and Centaurs on the outer planets and
their moons. For that purpose, we repeat the simulations in N17.
Specifically, we select the best case from N17,\footnote{This is the 
case that best matches various observational constraints, including the 
orbital structure of the Kuiper belt, number and orbital distribution 
of ECs, etc. Planet Nine was not included in the selected model (N17).
See N17 and Nesvorn\'y et al. (2020) for the range of Neptune migration 
histories that have been explored and for how various observational 
constraints were used to narrow the range of possibilities.} 
in which Neptune started at $a_{\rm N,0}=22$ au, migrated at a rate proportional to
$\exp(-t/\tau_1)$, with $\tau_1 = 10$~Myr, for the first 10~Myr, and
then at a rate $\propto \exp(-t/\tau_2)$, with $\tau_2=30$ Myr, for
$10<t<500$~Myr (Case 2 in N17). Time $t$ is measured from the
dispersal of the protoplanetary gas disk some 4.56 Gyr ago.  The
dynamical instability happened at $t=10$ Myr in this case.  The
present properties of the EC population are not sensitive to the
details of Neptune's early migration (e.g., Nesvorn\'y et al. 2017).  
The galactic tide and stellar encounters were included in the simulation, 
while the putative Planet Nine (Trujillo \& Sheppard 2014, Batygin 
\& Brown 2016, N17)  was not.
 
Since we are interested in the {\it current} impact rate of
ECs/Centaurs (see \citet{Wong2019}, \citet{Wong2021} and
\citet{Bottke2023} for impact rates in the early Solar System), we
repeated the last segment of the N17 simulation, in which cometary
reservoirs were evolved from 1 Gyr ago to the present time.  By
slicing this wide time interval into smaller segments and comparing
the results, we verified that the EC/Centaur population changed little
over the past 1 Gyr (e.g., the impact rate on Jupiter changes by
$<10$\% in the past 1 Gyr). We thus used the full statistics in the
1-Gyr long interval to infer the current impact rate. The original
simulations started with $10^6$ test planetesimals in the disk between
24 and 30 au at $t=0$. To further improve the model statistics, we
cloned test bodies as they evolved toward the inner Solar
System. Specifically, in the last 1 Gyr interval, we monitored the
heliocentric distance $r$ of each body, and cloned it 50 times when
$r$ first dropped below $r^*=23$ au.\footnote{The cloning was done 
by a small change of the velocity vector ($10^{-6}$ relative to the 
vector magnitude). We cloned at $r^*=23$ au because we wanted to 
have good statistics for Uranus impactors. Cloning at much larger
heliocentric distance was impractical because it would have generated 
excessive amount of data.} This effectively corresponds to
$N=5\times10^7$ (initial) test planetesimals. The simulations were
performed on 2000 Ivy Bridge cores of the NASA Ames Pleiades
Supercomputer. We used the \texttt{swift\_rmvs4} integrator
\citep{Levison1994} and a 0.2 yr timestep. All impacts of bodies on
the outer planets were recorded by the $N$-body integrator.
 
To compute impact rates on the outer planets' moons, we first recorded
all encounters of model bodies within $\simeq$1 Hill radius of each
planet. For every encounter, we then used Eq.~(3) from
\cite{Nesvorny2004} to compute the collisional probability with
moons.\footnote{See \citet{Kessler1981} and \citet{Nesvorny2003} for further
discussion. The main assumption of this method is that the orbital 
longitudes are randomly distributed between 0 and $2\pi$.} 
The computation of collisional probabilities accounts for the moons' real orbits, including their (typically small) orbital eccentricities and 
inclinations. We verified that, for a moon on a circular orbit, the
results are consistent with the \"Opik equations (\citet{Opik1951};
\citet{Zahnle1998}; \citet{Zahnle2003}; see footnote 8 in
\citet{Nesvorny2004}). The bodies that were bound to the planet were
treated separately -- their collisional probability was computed from
Eq.~(13) in \citet{Nesvorny2003}. In most cases, dynamical and
physical characteristics of moons (orbital elements, physical radii,
surface gravities, etc.) were taken from the NASA JPL Planetary
Satellites site.\footnote{\texttt{https://ssd.jpl.nasa.gov/sats/}} We
assumed spherical shapes for all moons and accounted for the
gravitational focusing of comets by moons. The reduction in outbound
impactor flux due to collisions with the planet (``shielding'') was
also accounted for \citep{Lissauer1988}.  We accumulated the
collisional probability over all recorded encounters and, following
Z03, expressed it as a fractional impact probability relative to
Jupiter (Tables 1--3).

We considered two ways an EC can become inactive: it becomes dormant,
either because a refractory mantle forms or the comet loses all
volatiles, or it breaks into small pieces and disappears. (We ignore
the possibility that the comet breaks into several large fragments.)
We must distinguish between these  possibilities because they
have different implications for the impact flux. The impact flux is
expected to be higher if comets become dormant, because the nucleus is
still intact.  If, instead, ECs disrupt, they must be removed from the
pool of impactors. Studies of meteor orbits suggest that disruption is
the main physical loss mechanism for Jupiter-family comets
\citep{Ye2016}.  Nonetheless, we considered two end-member models. In
each, we assumed that comets have a physical lifetime of $N_{\rm p}(2.5)=600$
perihelion passages within 2.5~au, as in N17 (Section 1).\footnote{N17 
studied several criteria for the physical lifetime of comets (time spent 
within 2.5 au of the Sun, a limit based on the accumulated insolation, etc.) and found 
that they produced similar results. The Kolmogorov-Smirnov (K-S) test was used 
to determine the best value of $N_{\rm p}(2.5)$ and its uncertainty. N17 found
the K-S test probability $>0.05$ for $N_{\rm p}(2.5)=300$--800. Here we choose 
$N_{\rm p}(2.5)=600$.} In the first
case (Sections 3.1.1 and 3.2.1), we assumed comets became dormant and
retained them in the impact flux calculation. In the second case
(Sections 3.1.2 and 3.2.2), we assumed comets were disrupted and
removed them from the impact flux calculation. We only model comet fading 
or disruption at {\it small} perihelion distances in this work (as parameterized 
by $N_p(2.5)$). Comet fading and/or disruption at larger perihelion distances 
(e.g., due to activity driven by supervolatiles or tidal encounters with planets) is not accounted for.

\section{Results}

\subsection{Planetary Impacts}

\subsubsection{Results Without Comet Disruption}

In total, our model recorded 217 direct impacts on Jupiter, 70 impacts
on Saturn and 62 impacts on Uranus. We do not consider impacts on
Neptune and its moons in this work. Neptune and its moons are
bombarded not only by Centaurs, but also by scattered disk objects
(SDOs). We do not have complete statistics for SDOs because we cloned
the bodies in our model at $r^*=23$ au. 
%(cloning at $r^*>30$ au is
%impractical because the volume of data would be excessive). 
Interestingly, 24\%, 11\% and 6\% of the bodies were bound
to Jupiter, Saturn and Uranus, respectively, at the time of 
impact.\footnote{Bound is defined as having negative total energy ($E$)
with respect to a planet or  impact speed lower than the escape speed at the planet's surface. 
We computed the potential and kinetic energies for all impactors at 
the time of impact. The impactors were then separated into bound 
($E<0$) and unbound ($E>0$) cases.}
For Jupiter, this is close to the 21\% reported in \citet{Levison2000}
and $15\pm2$\% given by \citet{Kary1996}. For Jupiter, 3 out of 217
impactors ($\sim 1.4$\%) had Tisserand parameters with respect to
Jupiter $T_{\rm J}<2$, indicating that only 1--2\% of impactors were
near-isotropic comets (NICs; LD97) from the Oort cloud
\citep{Vokrouhlicky2019}. The great majority of impactors (98--99\%)
were ECs with $2<T_{\rm J}<3.05$. Ecliptic comet precursors -- the
low-inclination Centaurs evolving from the scattered disk -- also
dominate impacts on Saturn and Uranus (e.g., only one of 62 Uranus
impactors had a retrograde heliocentric orbit).
%with $i>90^\circ$).

The impact rate of ECs/Centaurs on Jupiter is computed as
follows. Having effectively $5\times10^7$ test bodies in the original
trans-Neptunian disk ($10^6$ original planetesimals $\times$ 50 clones
for bodies that reached $r^*=23$ in the past billion years, see
Section 2), we recorded 217 impacts in 1 Gyr.  According to the
calibration discussed above, there were $(8 \pm 3) \times 10^9$
planetesimals with $D>10$ km when Neptune began to migrate. The
current rate of impacts of $D>10$ km bodies on Jupiter (i.e., the
average for the last billion years, see Section~2) is therefore $217
\times 8 \times 10^9 / (5 \times 10^7 \times 10^9) = 3.5 \times
10^{-5}$ yr$^{-1}$, implying a timescale for impacts by $D>10$-km
bodies $\simeq 2.9 \times 10^4$ yr.  If we adopt the reference
cumulative size distribution $N(D) \propto D^{-2.1}$ from N17 (Section
2) and assume it extends down to $D = 1$~km, we estimate a timescale
for impacts by $D>1$-km bodies of $\simeq 230$ yr. We use this
timescale as a baseline in the rest of this paper. For comparison,
\citet{Dones2009} inferred the impact flux on Jupiter from the
historical record of close approaches and impacts
\citep{Schenk2007}. They estimated an impact rate of at least $4
\times 10^{-3}$ yr$^{-1}$ for $D>1$~km, implying a timescale $\lesssim
250$ yr, which is consistent with our model results.  Saturn and
Uranus receive 0.32 and 0.29 of the Jupiter impact flux, respectively,
implying impact timescales of $\simeq 720$ yr and $\simeq 790$ yr,
respectively, for $D>1$-km impacts.

One major source of uncertainty of our model estimates is the
extrapolation of the impact rate from $D>10$ km to $D>1$ km. Above, we
assumed a single power-law slope of 2.1 for simplicity. Subaru telescope
observations of Jupiter Trojans indicate that the cumulative power
index of Jupiter Trojans changes from $\simeq 2.25$ for $D>5$ km to
$\simeq 1.8$ for $D<5$ km, assuming that albedo is independent of size
(e.g., \citet{Uehata2022}). If the shallower slope is extended all the
way down to 1 km, we would infer a $\simeq 340$ yr timescale for
impacts of $D>1$ km bodies on Jupiter, i.e., an impact rate $\sim 2/3$
of our previous estimate\footnote{If we assume $N(D) \propto
D^{-2.25}$ for diameters in the range 5--10 km, and $N(D) \propto
D^{-1.8}$ for diameters between 1 and 5~km, the impact rate is lower
than our original estimate, which assumes $N(D) \propto D^{-2.1}$
between 1 and 10~km, by a factor $(10/5)^{2.25}
(5/1)^{1.8}/(10/1)^{2.1} \approx 0.68$.}.  It also may be that the
size distribution of Jupiter Trojans is not a good proxy for cometary
impactors. If we generously assume that the cometary size distribution
is steeper, $N(D) \propto D^{-2.4}$, for $1<D<10$ km, based on craters
on Iapetus's dark terrain \citep{Kirchoff2010}, we obtain a $\simeq
120$-yr timescale for impacts of $D>1$ km bodies on Jupiter.
%In summary, our model estimate has a factor of $\sim 2$ uncertainty.

\subsubsection{Results With Comet Disruption}

\begin{figure}
%\epsscale{1.5}
\epsscale{1.5}
\hspace*{-2.cm}
\plotone{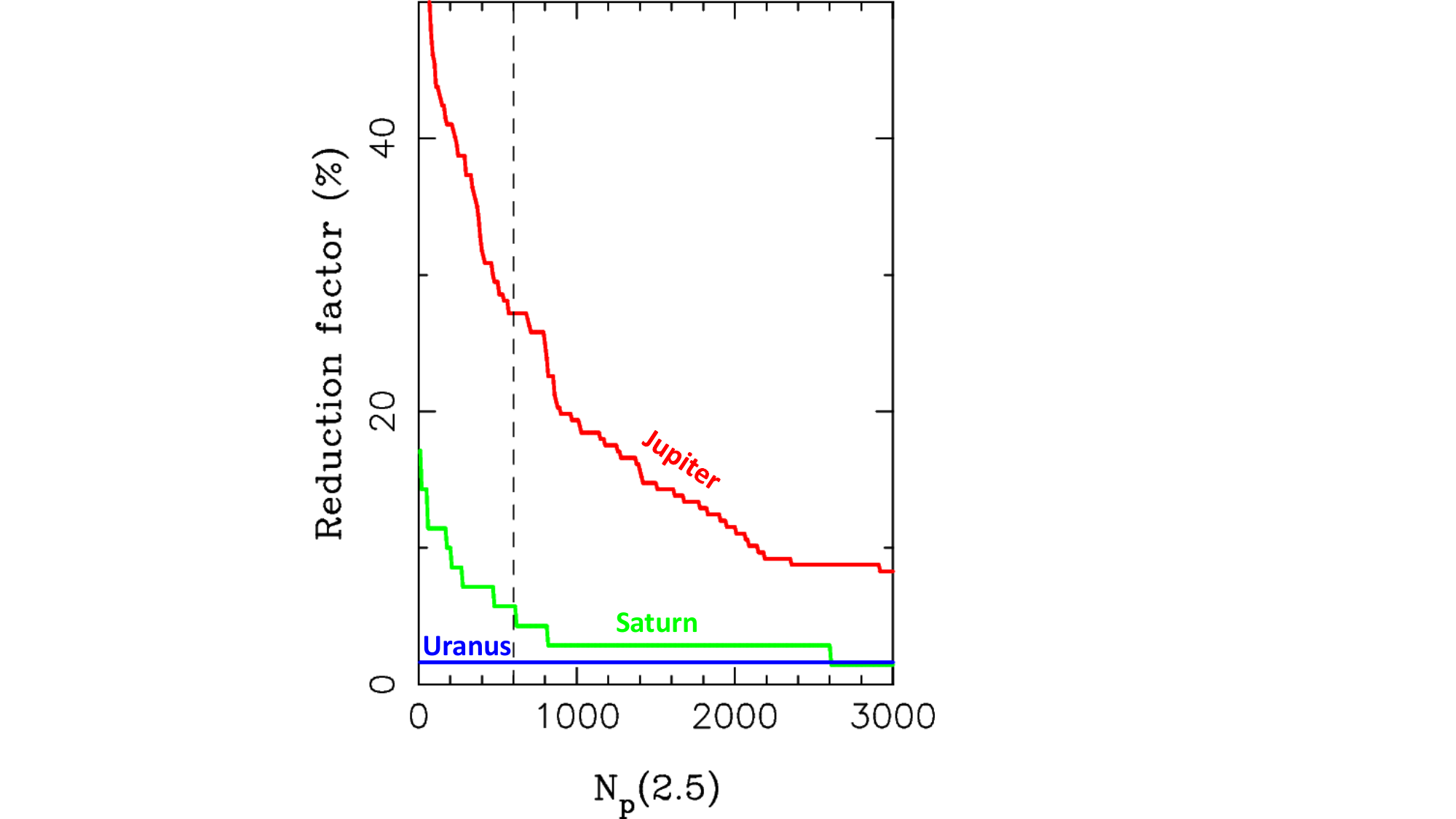}  
%\plotone{reduct.pdf}   
%\hspace*{-2.cm}\plotone{fig3.pdf}
\caption{The number of planetary impacts is reduced when comets are
  assumed to disrupt, on average, after $N_p(2.5)$ orbits with perihelion distances $q<2.5$
  au. To make this figure, we monitored the number of perihelion passages with $q<2.5$ au and 
discarded bodies for which the number of perihelion passages, $N_{\rm p}(2.5)$, exceeded the 
threshold given on the $X$ axis. We then computed the number of impacts from the remaining bodies 
and normalized it to the number of impacts for the case without comet disruption. The reduction
factor is plotted on the $Y$ axis.
  The results for Jupiter, Saturn and Uranus are shown by red,
  green and blue lines, respectively. For example, for $N_{\rm p}(2.5)
  = 600$ and Jupiter impacts, the impact rate is reduced by 27\%
  relative to its value neglecting disruption. }
\label{phys}
\end{figure}

Figure \ref{phys} shows how the number of planetary impacts is reduced
when we assume that comets disrupt after a certain number of orbits
with $q<2.5$ au. As expected, comet disruption has the largest effect
for Jupiter, where the impact rate, assuming $N_{\rm p}(2.5) = 600$,
is reduced by 27\% from its value neglecting disruption. For Saturn
and Uranus, the corresponding factors are 6\% and 2\%.\footnote{We remind
the reader that we only model comet disruption at small perihelion distances 
in this work, as parameterized by $N_{\rm p}(2.5)$ (e.g., tidal disruption 
of comets during planetary encounters is not modeled).
It is expected that that comet disruption at small perihelion distances should  
affect Jupiter impactors more than Saturn/Uranus impactors, because Jupiter 
impactors, given the smaller orbital radius of Jupiter, are more likely 
to evolve below 2.5 au before they can impact.}
The average time between impacts of $D>1$-km bodies on Jupiter, Saturn and Uranus
is then 320, 760 and 810 yr, respectively (here and elsewhere in the
paper we scale from the nominal reference timescale of 230~yr for
Jupiter discussed above). For all planets, the reduction factor
decreases if we assume long physical lifetimes for comets. For
example, for $N_{\rm p}(2.5)=3000$, as appropriate for $D\sim10$ km
ECs from N17, the impact flux on Jupiter is reduced by only 8\%.

The fraction of bound impactors is larger when comet disruption is
included in the model.  For example, without accounting for comet
disruption, we find that 24\% of Jupiter impactors were bound to
Jupiter when they impacted. With $N_{\rm p}(2.5)=600$, which should be
appropriate for $D \sim 1$ km comets (N17), we find that the fraction
of bound impactors is $\simeq30$\%.  Bound Jupiter impactors are
slightly less likely to have many perihelion passages below 2.5 au
than the population of ECs as a whole. Bound objects typically have
$T_{\rm J} \approx 3$ \citep{Kary1996}.  Bodies that encounter Jupiter
at higher velocities, and hence have smaller Tisserand parameters, are
more easily scattered by Jupiter into orbits with small perihelion
distances (\citet{Levison1997}; see Figure~9 of
\citet{Fernandez2018}). This means that events such as Shoemaker-Levy
9, which was tidally disrupted by Jupiter in 1992 and collided with
the planet in 1994, would happen slightly more often -- relative to
impacts from unbound orbits -- if comets disrupt.\footnote{Here we only 
model comet disruption at low perihelion distances. When we say 
that ``events such as Shoemaker-Levy
9 ... would happen slightly more often," we mean 
that there should be relatively more impacts on Jupiter where the 
impactor is bound to Jupiter prior to an impact. That includes all cases for which the impactor's orbital energy with
respect to Jupiter is negative, whether or not the impactor was tidally
disrupted on a previous orbit. Note that we do not model 
tidal disruption, and so cannot say what the fraction 
of tidally disrupted impactors should be in different cases.}

\subsection{Impacts on Moons}

\subsubsection{Results Without Comet Disruption}
 
\begin{figure}
\epsscale{1.0}
%\hspace*{1.cm}\plotone{velo.pdf}
%\plotone{velo2.jpg}
\plotone{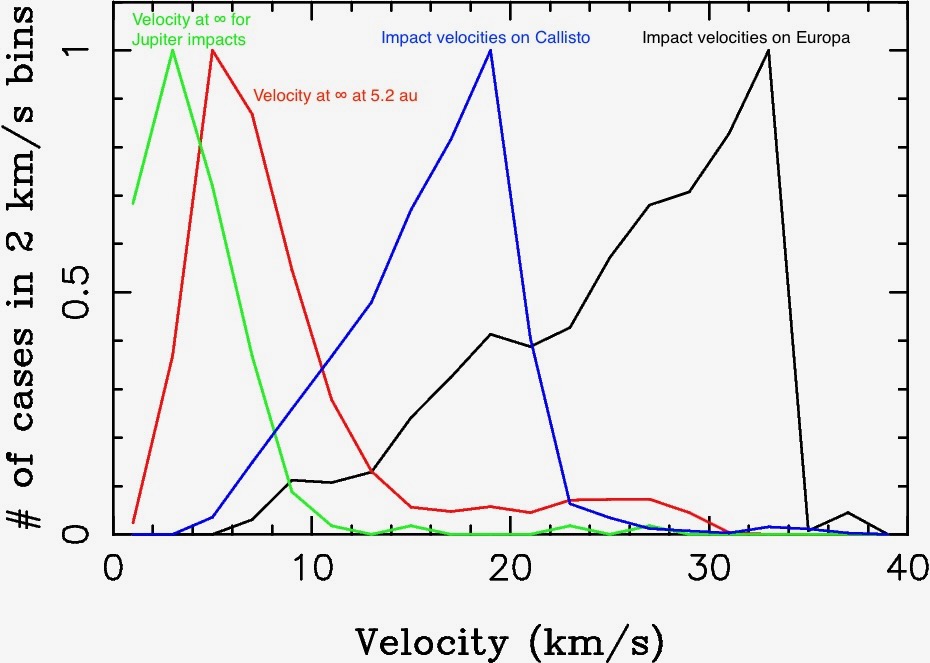}
\caption{The velocity distributions inferred in this work (all
  normalized to 1). The red line shows the velocity distribution for
  comets impacting a small target at the orbital distance 5.2 au from
  the Sun -- the effects of gravitational focusing are neglected in
  this case. The green line is the velocity ``at infinity'',
  $v_\infty$, for comets impacting Jupiter. The velocities are smaller
  in this case because Jupiter impacts preferentially sample the
  low-velocity tail of the background distribution (bodies with small
  encounter velocities are strongly focused and have larger impact
  probabilities). The blue and black lines are the impact velocities
  of comets on Europa and Callisto, respectively. The impact
  velocities are larger because they include the contribution from the
  orbital motion of moons and gravitational focusing.}
\label{velo}
\end{figure}

The impact probabilities on the outer planets' moons, both with and
without comet disruption, are reported in Tables 1--3. There are some
notable differences from \citet{Zahnle2003}. We obtain higher impact
speeds for the outermost moons. For example, we find $\langle v_{\rm
  imp} \rangle = 7.1$ km/s for Phoebe, whereas Z03 reported $\langle
v_{\rm imp} \rangle = 3.2$ km/s (here, $v_{\rm imp}$ is the impact speed
of an individual body and $\langle v_{\rm imp} \rangle$ is the mean 
impact speed computed over all recorded impacts). Z03 ultimately based 
their encounter velocity ($v_\infty$, the velocity at ``infinity'' -- in practice, the planet's Hill sphere) distribution on objects that struck Jupiter in
LD97. That distribution is biased toward low-velocity encounters
because of gravitational focusing by Jupiter. In reality, there is a
wide range of values of $v_\infty$ at each planet
(Fig.~\ref{velo}). The outermost satellites of the giant planets have
orbital velocities $v_{\rm orb} < \langle v_\infty \rangle$, where
$\langle v_\infty \rangle$ is the average encounter velocity of bodies
at the planet's Hill sphere ($v_{\rm orb}$ is the orbital velocity of a moon
around its parent planet). Therefore, gravitational focusing by the
planet is a minor effect for them. This implies that impacts on the
outer moons sample the high-velocity portion of the background
velocity distribution more than do the inner moons and the planets
themselves (\citet{Zahnle1998}, Section 2.1).

By contrast, the inner satellites of the giant planets have orbital
velocities $v_{\rm orb} > \langle v_\infty \rangle$. As a result,
Centaurs/ECs are more strongly focused gravitationally. In the limit
$v_{\rm orb} \gg \langle v_\infty \rangle$, the impact rate per unit
area on a moon ${\cal F}_{\rm imp} \appropto 1/(a_{\rm moon}
v_\infty^2$), where $a_{\rm moon}$ is the moon's semimajor axis (see,
e.g., Eq.~(4) in Z03).  For reference, $\langle v_\infty \rangle =
$~4.1~km/s for bodies that impact Jupiter, while $\langle v_\infty
\rangle = $~8.8~km/s for bodies that cross Jupiter's orbit. These
values are, respectively, approximately 30\% and 70\% of Jupiter's
mean orbital speed.

Other differences may arise from the dependence on the structure of
the source regions assumed in different works. Z03 adopted the results
from \citet{Levison1997, Levison2000}, where bodies started with low
inclinations in the classical Kuiper belt. This leads to a dynamically
colder population of Centaur/EC impactors and lower impact speeds for
more distant satellites. Here, instead, Centaurs/ECs evolve from the
scattered disk in our model (N17) and so have higher inclinations and
larger impact speeds.

The mean impact speeds for the inner moons that we find in this work
are typically slightly smaller than the mean impact speeds given in
Z03. For a monodisperse encounter velocity ($v_\infty$), higher impact
speeds for the outer moons would imply higher impact speeds for the
inner moons as well.  If we approximate the distribution of $v_\infty$
as a Maxwellian, the impact probability weighted mean encounter
velocity is $\frac{1}{2} v_\infty$ in the limit $v_{\rm orb} \gg
\langle v_\infty \rangle$. The slightly lower impact speeds for the
inner moons probably reflect this sampling and the different velocity
distributions we and Z03 assume (see \citet{Zahnle2001}, Eq.~1).  Our
low-velocity tail probably extends to slightly lower speeds than does
that of Z03, and this tail is preferentially sampled by the inner
moons.

We find higher impact rates for distant moons and lower impact rates
for the innermost moons (Tables 1--3; recall that these rates are
normalized to Jupiter), relative to Z03. Therefore, our impact
probability profiles with orbital radius are flatter than in Z03. For
example, the outer moons of Saturn have impact probabilities that are
up to $\sim 2$ times higher than in Z03 (e.g., Phoebe has $P_{\rm
  imp}=1.8\times 10^{-8}$ here and $P_{\rm imp}=8.7\times 10^{-9}$ in
Z03). The higher (normalized) impact probabilities for the outermost
moons are a consequence of higher encounter velocities in our model,
which result in lower gravitational focusing on the planets.
%The larger $\langle v_\infty \rangle$ must
%also be the cause of the slightly smaller impact rates on the inner
%moons, but we have not yet investigated this issue in detail.

The mid-sized moons of Saturn have impact probabilities that are
slightly smaller than reported in Z03 (e.g., Mimas has $P_{\rm
  imp}=1.7\times 10^{-6}$ in Z03 and $P_{\rm imp}=1.1\times 10^{-6}$
in this work; the impact probability $P_{\rm imp}$ is relative to
Jupiter). Part of this difference results from the lower impact rates
on Saturn relative to Jupiter that we find here (0.32 vs.\ 0.42 in Z03).

For the innermost moons (Metis--Thebe for Jupiter, Pan--Janus for
Saturn, and Portia and Puck for Uranus), the differences with Z03 are
larger. These result from, e.g., differences in assumed sizes for the
satellites and shielding by the planet \citep{Lissauer1988}, which we
account for, but Z03 did not. As an independent check, we verified
that our impact probabilities for the innermost moons follow the
expected scaling.  For these moons, the value of $P_{\rm imp}$ should
be approximately $(R_{\rm moon}/R_{\rm planet})^2 (R_{\rm planet}/a)$,
where $R_{\rm moon}$ and $R_{\rm planet}$ are the physical radii of
the moon and planet, and $a$ is the semi-major axis of the moon. The
first term is the ratio of cross sections, while the second term
accounts for gravitational focusing by the planet.

\subsubsection{Results With Comet Disruption}

The impact flux is reduced when the effects of comet disruption are
included in the model (Tables 1--3).  The flux is reduced by $\sim
25$--45\% for the jovian moons, $\sim 5$--15\% for the saturnian
moons, and $<1$\% for the uranian moons. This trend, with the
reduction factor decreasing with heliocentric distance, is expected
because comet disruption mainly reduces the impactor population at
smaller heliocentric distances. Compared to planetary impacts (Section
3.1), the reduction factors for moons are larger. For example, the
impact flux on Jupiter was reduced by 27\% from the original flux for
$N_{\rm p}(2.5)=600$ (appropriate for $\sim 1$-km impactors).  For
jovian moons, however, the impact flux is reduced by $\sim 25$--45\%
for $N_{\rm p}(2.5)=600$, and there is a trend with smaller reduction
for inner moons and larger reduction for outer moons.  

The impact speeds on the outer moons of Jupiter are slightly lower in
the case with disruption. For example, the mean impact speeds on
Himalia are 8 and 9~km/s with and without disruption, respectively
(Table~1). This most likely happens because in the case with
disruption, there is not enough time to excite the heliocentric orbits of EC impactors as much. The
effects of disruption on the impact speeds for moons of Saturn and
Uranus are small (Tables 2 and 3).

The lower cratering rates in the case that includes cometary
disruption imply somewhat older surface ages for lightly-cratered
terrains on outer planet satellites such as Enceladus and Europa.  For
example, Z03 estimated the surface age of Europa between 30 and 70 Myr
(more recent papers infer slightly older ages: 60--100~Myr in
\citet{Zahnle2008} and 40--90~Myr in \citet{Bierhaus2009}). Scaling
from Z03, our lower impact flux in the model with comet disruption
would imply a surface age between 45 and 105 Myr. 
%{\bf David: I think Bill Bottke is getting about 300~Myr for Europa.} 
The catastrophic
disruption timescales of small moons reported in Z03, which assume
that a crater with a diameter equal to the diameter of the satellite
dooms the moon, would also be longer.

\begin{figure}
%\epsscale{0.6}
%\plotone{nprofile.eps}
%\plotone{wprofile.eps}
%\epsscale{2.0}
\epsscale{1.5}
%\hspace*{-9.cm}\plotone{fig5.pdf}
%\vspace{-4in}
\hspace*{-4cm}
%\plotone{fig5-rev.pdf}
\plotone{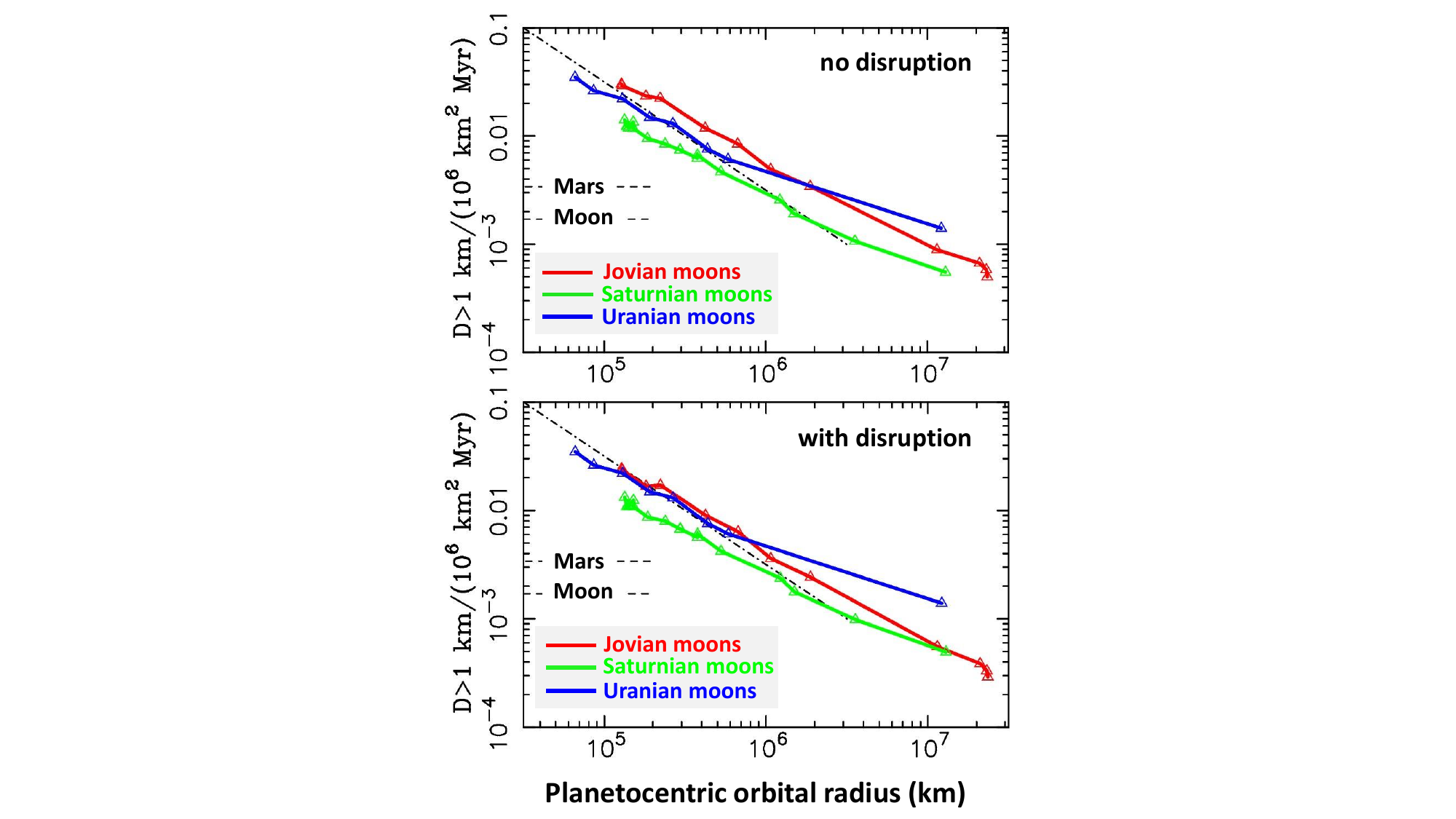}
\caption{The impact flux of $D>1$ km comets on jovian (red line and
  triangles), saturnian (green) and uranian (blue) moons .  The flux
  is given per $10^6$ km$^2$ per Myr. For reference, the horizontal
  dashed lines show the asteroid impact flux on the Moon and Mars,
  both for $D>1$ km bodies per $10^6$ km$^2$ of surface per Myr
  (Nesvorn\'y et al. 2022).  The top panel disregards comet
  disruption, while the bottom panel includes comet disruption with
  $N_{\rm p}(2.5)=600$. The dashed-dotted line is ${\cal F}_{\rm imp}
  \propto 1/a_{\rm moon}$, plotted here for reference.}
\label{prof}
\end{figure}

Figure \ref{prof} shows how the impact flux depends on the orbital
radius of a moon. Here, we disregard the moons' sizes by normalizing
the impact flux of $D>1$ km comets per $10^6$ km$^2$ of the moon's
surface (per Myr).  When comet disruption is not accounted for, the
normalized impact flux is ${\cal F}_{\rm imp}
\simeq$~0.0005--0.002/$(10^6 \, {\rm km}^2 \, {\rm Myr})$ for the
outer moons and ${\cal F}_{\rm imp} \simeq 0.01$--0.05/$(10^6 \, {\rm
  km}^2 \, {\rm Myr})$ for the inner moons. The higher flux on inner
moons is due to gravitational focusing by the planets.
This trend persists when we account for comet disruption (bottom panel
of Fig.~\ref{prof}).  Given that the flux reduction is larger for the
outer moons, however, the overall slope of the ${\cal F}_{\rm imp}$
dependence on $a_{\rm moon}$ becomes slightly steeper with comet
disruption. For example, the outer/irregular moons of Jupiter have
${\cal F}_{\rm imp} \simeq$~0.0004/$(10^6 \, {\rm km}^2 \, {\rm Myr})$
with $N_{\rm p}(2.5)=600$ compared to ${\cal F}_{\rm imp}
\simeq$~0.0006/$(10^6 \, {\rm km}^2 \, {\rm Myr})$ without comet
disruption.

\section{Discussion}

The impact flux on the satellites depends on whether comets undergo
disruption or become dormant. The flux is higher if comets become
dormant (Tables 1--3). We expect that both cases apply to some
degree. For example, if all comets become dormant, the ratio of the
number of dormant to active ECs should be roughly 30--60. This is
because the dynamical lifetime of ECs with $q < 2.5$~au is $t_{\rm
  dyn} \sim 3 \times 10^5$~yr (LD97), while their physical lifetime is
$t_{\rm phy} \sim$~5,000--12,000~yr (LD97, N17). The ratio of the
number of dormant to active ECs should be $\sim t_{\rm dyn}/t_{\rm
  phys}$.

\citet{Licandro2016} obtained WISE observations of asteroids on
(mostly Jupiter-family-like) cometary orbits [ACOs] and inferred their
size distribution. They found that the number of ACOs is smaller than
the number of JFCs for diameters $< 10$~km. This suggests that most
ecliptic comets end their lives by being disrupted instead of becoming
dormant (also see \citet{Ye2016}). Therefore, the impact fluxes that
we calculate for the case with cometary disruption are likely to be
more realistic. This may not be true for Centaurs, which are less
likely to be disrupted because of their greater distances from the Sun.

\section{Conclusions}

The main results of this work can be summarized as follows:

\begin{enumerate}

\item We determined the current rates at which comets and Centaurs
  strike the moons of Jupiter, Saturn, and Uranus.  Compared to
  \citet{Zahnle2003}, we find a higher impact flux on the outer moons
  and a smaller impact flux on the inner moons. This is a consequence
  of the larger orbital inclinations in our model, in which the
  scattered disk is the main source of comets/Centaurs. The impact
  speeds on outer moons are significantly higher (e.g., 7.1 km/s for
  Phoebe compared to 3.2 km/s reported for Phoebe in Z03), implying
  larger craters for a given impactor size.

\item When comet disruption is accounted for, the impact flux on the
  outer planets and their moons is reduced. The reduction factor
  depends on the adopted disruption model. For example, for $N_{\rm
    p}(2.5)=600$, which was the preferred value for $D \sim 1$ km
  comets in \citet{Nesvorny2017}, the impact flux on Jupiter is
  reduced to 73\% of the original value (the mean interval between
  impacts of $D>1$ km comets on Jupiter is $\sim 230$ yr without comet
  disruption and $\sim 315$ yr with comet disruption). 

\item For $N_{\rm p}(2.5)=600$, the impact flux is reduced by $\sim 25$--45\%
  for jovian moons, $\sim 5$--15\% for saturnian moons, and $<1$\% for uranian
  moons. The lower cratering rates in the
  case that includes cometary disruption implies somewhat older
  surface ages for lightly-cratered terrains on outer planet
  satellites such as Enceladus and Europa.  For example, the reduced
  impact fluxes with comet disruption would imply older surfaces
  (e.g., a $\sim 45$--105 Myr surface age for Europa).

\item The impact rate on Jupiter changes by less than 10\% in the past
  1 Gyr. \citet{Wong2019} find a much stronger decay during the first
  billion years of the Solar System. Together, these results show that
  most of the cratering occurred early on \citep{Wong2021,
  Bottke2023}.

\item The average time between impacts of $D>1$ km bodies, for the case
    with comet disruption and our nominal size distribution, is 
    2.7 Myr, 5.1 Myr, 3.2 Myr and 5.7 Myr for Io, Europa, Ganymede and 
    Callisto, respectively (Table 1). It is 42 Myr, 45 Myr, 32 Myr and 5.0 Myr
    for Tethys, Dione, Rhea and Titan (Table 2), and 16 Myr, 18 Myr, 17 Myr and
    23 Myr for Ariel, Umbriel, Titania and Oberon (Table 3).   
\end{enumerate}

\acknowledgements

The simulations were performed on the NASA Pleiades Supercomputer. We
thank the NASA NAS computing division for continued support. The work
of DN was supported by the NASA Solar System Workings program. The
work of DN, LD, and MDP was supported by the NASA Cassini Data
Analysis Program. This material is based in part on work done by MW
while serving at the National Science Foundation. We thank the reviewers,
Darryl Seligman and Wes Fraser, for helpful suggestions to the
submitted manuscript.

\clearpage

\begin{table}
\centering
{
\begin{tabular}{lrrrrrrr}
\hline \hline & \multicolumn{2}{c}{\it Zahnle et al.\ (2003)} &
\multicolumn{5}{c}{\it This Work} \\ & & & \multicolumn{2}{c}{\it No
  Disruption} & \multicolumn{3}{c}{\it With Disruption} \\ & $P_{\rm
  imp}$ & $\langle v_{\rm imp} \rangle$ & $P_{\rm imp}$ & $\langle
v_{\rm imp} \rangle$ & $P_{\rm imp}$ & $\langle v_{\rm imp} \rangle$ & $\tau_{\rm imp}$
\\ & & km/s & & km/s & & km/s & Myr \\ \hline {\bf Jupiter} & 1.0 & -- & 1.0 & --
& 0.73 & -- & -- \\
\hspace{3.mm}Metis & $2.8\times10^{-7}$ & 59 & $4.0\times10^{-8}$ &
58.2 & $3.2\times10^{-8}$ & 58.7 & -- \\
\hspace{3.mm}Adrastea & -- & -- & $5.6\times10^{-9}$ & 58.0 &
$4.5\times10^{-9}$ & 58.5 & --  \\
\hspace{3.mm}Amalthea & $7.7\times10^{-7}$ & 50 & $4.7\times10^{-7}$ &
46.2 & $3.3\times10^{-7}$ & 45.8 & 680 \\
\hspace{3.mm}Thebe & $2.9\times10^{-8}$ & 45 & $1.5\times10^{-7}$ &
43.2 & $1.2\times10^{-7}$ & 43.5 & 1900 \\
\hspace{3.mm}Io & $1.4\times10^{-4}$ & 32 & $1.1\times10^{-4}$ & 31.6
& $8.5\times10^{-5}$ & 31.5 & 2.7 \\
\hspace{3.mm}Europa & $6.6\times10^{-5}$ & 26 & $5.9\times10^{-5}$ &
25.4 & $4.4\times10^{-5}$ & 25.2 & 5.1 \\
\hspace{3.mm}Ganymede & $1.2\times10^{-4}$ & 20 & $9.7\times10^{-5}$ &
20.3 & $7.1\times10^{-5}$ & 20.0 & 3.2 \\
\hspace{3.mm}Callisto & $6.1\times10^{-5}$ & 15 & $5.7\times10^{-5}$ &
16.0 & $4.0\times10^{-5}$ & 15.5 & 5.7 \\
\hspace{3.mm}Himalia & $1.4\times10^{-8}$ & 6.1 & $1.8\times10^{-8}$ &
9.0 & $1.1\times10^{-8}$ & 8.0 & --  \\
\hspace{3.mm}Ananke & -- & -- & $4.0\times10^{-10}$ & 8.5 &
$2.3\times10^{-10}$ & 7.2 & -- \\
\hspace{3.mm}Carme & -- & -- & $9.0\times10^{-10}$ & 8.5 &
$5.2\times10^{-10}$ & 7.2 & -- \\
\hspace{3.mm}Pasiphae & -- & -- & $1.2\times10^{-9}$ & 8.4 &
$6.9\times10^{-10}$ & 7.2 & -- \\ \hline \hline
\end{tabular}
}
\caption{The impact probabilities $P_{\rm imp}$ and average impact
  speeds $\langle v_{\rm imp} \rangle$ for jovian moons.  The impact
  probabilities are given with respect to the impact probability on
  Jupiter, for the case with no comet disruption (Section 3.1.1). 
  To infer the impact flux on a moon, $P_{\rm imp}$ needs to be 
  multiplied by the impact rate of bodies on Jupiter.
  In Section 3.1.1, we infer a rate of $3.5 \times 10^{-5}$/yr for bodies
  with diameters $D > 10$~km striking Jupiter; the corresponding rate
  on Europa is $(5.9 \times 10^{-5}) (3.5 \times 10^{-5})$/yr, or $2.1
  \times 10^{-9}$/yr, which implies that a 10-km comet should strike
  Europa about once in 500~Myr at the present time. For our nominal
  size distribution, ($N(D) \propto D^{-2.1}$ for impactor diameters
  between 1 and 10~km (Section 2)), the rate at which 1-km comets
  strike Europa is greater by a factor $10^{2.1}$, implying a rate of
  $2.6 \times 10^{-7}$/yr and a timescale of $\sim 4$~Myr for impacts
  by km-sized bodies. If cometary disruption is considered, the impact
  rate on Europa is $(4.4 \times 10^{-5}) (3.5 \times 10^{-5})$/yr for
  10-km bodies, or $1.5 \times 10^{-9}$/yr, implying about once such
  impact in 700 Myr is expected. If we again assume $N(D)
  \propto D^{-2.1}$ for impactors between 1 and 10~km, the
  corresponding values for km-sized bodies striking Europa are 
  $1.9\times 10^{-7}$/yr and a timescale of $\sim 5$~Myr. The last 
  column reports the average time between impacts of $D>1$ km bodies ($\tau_{\rm imp}$),
  for the case with comet disruption and our nominal size 
  distribution (timescales exceeding the age of the Solar System
  are not shown).}
\label{tab1} 
\end{table}

\clearpage
\begin{table}
\centering
{
\begin{tabular}{lrrrrrrr}
\hline \hline & \multicolumn{2}{c}{\it Zahnle et al.\ (2003)} &
\multicolumn{5}{c}{\it This Work} \\ & & & \multicolumn{2}{c}{\it No
  Disruption} & \multicolumn{3}{c}{\it With Disruption} \\ & $P_{\rm
  imp}$ & $\langle v_{\rm imp} \rangle$ & $P_{\rm imp}$ & $\langle
v_{\rm imp} \rangle$ & $P_{\rm imp}$ & $\langle v_{\rm imp} \rangle$ & $\tau_{\rm imp}$
\\ & & km/s & & km/s & & km/s & Myr \\ \hline {\bf Saturn} & 0.42 & -- & 0.32 &
-- & 0.30 & -- & -- \\
\hspace{3.mm}Pan & -- & -- & $7.9\times10^{-9}$ & 30.1 &
 $7.4\times10^{-9}$ & 30.6 & -- \\
\hspace{3.mm}Atlas & -- & -- & $8.0\times10^{-9}$ & 28.5 &
 $7.0\times10^{-9}$ & 29.3 & -- \\
\hspace{3.mm}Prometheus & $1.7\times10^{-7}$ & 32 &
 $6.5\times10^{-8}$ & 29.6 & $5.9\times10^{-8}$ & 30.4 & 3800 \\
\hspace{3.mm}Pandora & $1.0\times10^{-7}$ & 31 & $5.6\times10^{-8}$ &
 29.6 & $5.1\times10^{-8}$ & 30.3 & 4400 \\
\hspace{3.mm}Epimetheus & $1.8\times10^{-7}$ & 30 & $1.3\times10^{-7}$
& 29.1 & $1.2\times10^{-7}$ & 30.0 & 1900 \\
\hspace{3.mm}Janus & $4.5\times10^{-7}$ & 30 & $2.7\times10^{-7}$ &
28.6 & $2.5\times10^{-7}$ & 29.4 & 920 \\
\hspace{3.mm}Mimas & $1.7\times10^{-6}$ & 27 & $1.1\times10^{-6}$ &
26.2 & $9.7\times10^{-7}$ & 26.6 & 230 \\
\hspace{3.mm}Enceladus & $2.2\times10^{-6}$ & 24 & $1.5\times10^{-6}$
& 23.1 & $1.4\times10^{-6}$ & 23.3 & 160 \\
\hspace{3.mm}Tethys & $7.9\times10^{-6}$ & 21 & $6.0\times10^{-6}$ &
21.0 & $5.4\times10^{-6}$ & 20.9 & 41 \\
\hspace{3.mm}Telesto & $3.5\times10^{-9}$ & 21 & $3.3\times10^{-9}$ &
21.0 & $3.0\times10^{-9}$ & 20.9 & -- \\
\hspace{3.mm}Calypso & $3.5\times10^{-9}$ & 21 & $2.4\times10^{-9}$ &
21.0 & $2.2\times10^{-9}$ & 20.9 & -- \\
\hspace{3.mm}Dione & $7.1\times10^{-6}$ & 19 & $5.6\times10^{-6}$ &
18.7 & $5.1\times10^{-6}$ & 18.4 & 45 \\
\hspace{3.mm}Helene & $5.8\times10^{-9}$ & 19 & $6.2\times10^{-6}$ &
18.7 & $5.6\times10^{-6}$ & 18.6 & -- \\
\hspace{3.mm}Rhea & $9.6\times10^{-6}$ & 16 & $7.8\times10^{-6}$ &
15.8 & $7.0\times10^{-6}$ & 15.8 & 32 \\
\hspace{3.mm}Titan & $5.4\times10^{-5}$ & 10.5 & $4.8\times10^{-5}$ &
11.2 & $4.5\times10^{-5}$ & 11.2 & 5.0 \\
\hspace{3.mm}Hyperion & $1.0\times10^{-7}$ & 9.4 & $1.0\times10^{-7}$
& 10.4 & $9.2\times10^{-8}$ & 10.3 & 2500 \\
\hspace{3.mm}Iapetus & $1.4\times10^{-6}$ & 6.1 & $1.6\times10^{-6}$ &
7.9 & $1.5\times10^{-6}$ & 7.7 & 150 \\
\hspace{3.mm}Phoebe & $8.7\times10^{-9}$ & 3.2 & $1.8\times10^{-8}$ &
7.1 & $1.6\times10^{-8}$ & 7.0 & -- \\ \hline \hline
\end{tabular}
}
\caption{The impact probabilities $P_{\rm imp}$ and average impact
  speeds $\langle v_{\rm imp} \rangle$ for saturnian moons. The impact
  probabilities are given with respect to the impact probability on
  Jupiter, for the case with no comet disruption. To infer the rate 
  of impacts on a moon, $P_{\rm imp}$ needs to be multiplied by the 
  impact rate of bodies on Jupiter (see Sect. 3.1.1 and the caption of Table 1). 
  For example, the rate of impact of 1-km bodies on Enceladus is $6.7 \times
  10^{-9}$/yr, or one such impact every $\sim 150$~Myr, for our
  nominal size distribution and no cometary disruption. For the same
  size distribution, but with cometary disruption, the corresponding
  rate is $6.4 \times 10^{-9}$/yr and the timescale of $\sim
  160$~Myr. The last column reports the average time between impacts of 
  $D>1$ km bodies ($\tau_{\rm imp}$), for the case with comet disruption and our nominal size 
  distribution (timescales exceeding the age of the Solar System
  are not shown).}
\end{table}

\clearpage

\begin{table}
\centering
{
\begin{tabular}{lrrrrrrr}
\hline \hline & \multicolumn{2}{c}{\it Zahnle et al.\ (2003)} &
\multicolumn{5}{c}{\it This Work} \\ & & & \multicolumn{2}{c}{\it No
  Disruption} & \multicolumn{3}{c}{\it With Disruption} \\ & $P_{\rm
  imp}$ & $\langle v_{\rm imp} \rangle$ & $P_{\rm imp}$ & $\langle
v_{\rm imp} \rangle$ & $P_{\rm imp}$ & $\langle v_{\rm imp} \rangle$ & $\tau_{\rm imp}$
\\ & & km/s & & km/s & & km/s & Myr \\ \hline {\bf Uranus} & 0.25 & -- & 0.29 &
-- & 0.28 & -- & -- \\
\hspace{3.mm}Portia & $4.7\times10^{-7}$ & 18 & $4.5\times10^{-7}$ &
16.9 & $4.5\times10^{-7}$ & 16.9 & 500 \\
\hspace{3.mm}Puck & $6.6\times10^{-7}$ & 15 & $4.9\times10^{-7}$ &
14.1 & $4.9\times10^{-7}$ & 14.1 & 460 \\
\hspace{3.mm}Miranda & $5.2\times10^{-6}$ & 12.5 & $3.5\times10^{-6}$
& 12.7 & $3.5\times10^{-6}$ & 12.7 & 65 \\
\hspace{3.mm}Ariel & $2.1\times10^{-5}$ & 10.3 & $1.4\times10^{-5}$ &
10.6 & $1.4\times10^{-5}$ & 10.6 & 16 \\
\hspace{3.mm}Umbriel & $1.6\times10^{-5}$ & 8.7 & $1.3\times10^{-5}$ &
9.2 & $1.3\times10^{-5}$ & 9.2 & 18 \\
\hspace{3.mm}Titania & $1.8\times10^{-5}$ & 6.8 & $1.3\times10^{-5}$ &
7.3 & $1.3\times10^{-5}$ & 7.3 & 17 \\
\hspace{3.mm}Oberon & $1.3\times10^{-5}$ & 5.9 & $1.0\times10^{-5}$ &
6.5 & $1.0\times10^{-5}$ & 6.5 & 23 \\
\hspace{3.mm}Sycorax & -- & -- & $2.6\times10^{-8}$ & 3.9 &
$2.5\times10^{-8}$ & 3.9 & -- \\ \hline \hline
\end{tabular}
}
\caption{The impact probabilities $P_{\rm imp}$ and average impact
  speeds $\langle v_{\rm imp} \rangle$ for uranian moons. The impact
  probabilities are given with respect to the impact probability on
  Jupiter, for the case with no comet disruption. To infer the rate 
  of impacts on a moon, $P_{\rm imp}$ needs to be multiplied by the 
  impact rate of bodies on Jupiter (see Sect. 3.1.1 and the caption of 
  Table 1). For example, the rate of impact of 1-km bodies on Miranda is 
  $1.5 \times 10^{-8}$/yr, or one such impact every $\sim 65$~Myr, for 
  our nominal size distribution (the results with and without cometary
  disruption are the same). The last 
  column reports the average time between impacts of $D>1$ km bodies ($\tau_{\rm imp}$),
  for the case with comet disruption and our nominal size 
  distribution (timescales exceeding the age of the Solar System
  are not shown).}
\end{table}

\clearpage

\clearpage

\clearpage

\clearpage

\clearpage

\end{document}